\documentclass{article}
\usepackage{spconf,amsmath,amssymb,graphicx,booktabs,multirow,placeins}
\usepackage[hidelinks]{hyperref}
\usepackage{cite}
\usepackage{etoolbox}
\patchcmd{\thebibliography}{\section{References}}
  {{\normalsize\section{References}}}{}
  {\PackageError{SPATEX}{Bibliography heading patch failed}{Check the spconf bibliography definition.}}
\apptocmd{\thebibliography}{%
  \setlength{\baselineskip}{9.7pt}%
  \setlength{\itemsep}{0pt}%
  \setlength{\parsep}{0pt}%
  \setlength{\parskip}{0pt}}{}
  {\PackageError{SPATEX}{Bibliography spacing patch failed}{Check the bibliography definition.}}
\usepackage{enumitem}
\setlist[itemize]{topsep=2pt plus .5pt minus .5pt, partopsep=0pt, itemsep=1pt, parsep=0pt}

\newcommand{\sisnri}{SI-SNRi}
\newcommand{\snri}{SNRi}
\newcommand{\dipd}{$\Delta$IPD}
\newcommand{\ditd}{$\Delta$ITD}
\newcommand{\dild}{$\Delta$ILD}
\newcommand{\doadrift}{DOA drift}
\newcommand{\presfive}{Pres@5}

\newcommand{\bibfont}{\normalsize}

\title{SPATEX: Array-Flexible Multichannel-to-Multichannel Target Speech Extraction using Direction Clue}

\name{\shortstack{Wen Wen$^{1,5,*}$, Changda Chen$^{2,*}$, Yu Xi$^{1,5}$, Haoyu Li$^{3}$, Qiang Zhou$^{1,4}$, Guanyu Chen$^{1,5}$, \\Xiaoyu Gu$^{1,5}$, Bohan Li$^{1,5}$, Shuai Wang$^{3}$, and Kai Yu$^{1,5,\dagger}$}
\thanks{\fontsize{9}{11}\selectfont 
\textsuperscript{*}denotes equal contribution.\newline
\textsuperscript{$\dagger$} denotes the corresponding author.}
}
\address{\parbox{0.97\textwidth}{\centering\fontsize{9}{11}\selectfont
$^{1}$ MoE Key Lab of Artificial Intelligence, AI Institute, X-LANCE Lab, Shanghai Jiao Tong University, China\\
$^{2}$ Waseda University, Japan \qquad $^{3}$ Nanjing University\qquad
$^{4}$ AISpeech Ltd, Suzhou, China\\
$^{5}$ Jiangsu Key Lab of Language Computing, Suzhou, China\\
\texttt{wenyideng@sjtu.edu.cn, kai.yu@sjtu.edu.cn}}}

\makeatletter
\def\@maketitle{\newpage
 \null
 \vspace{4pt}\begin{center}
 {\large\bf\@title\par}\vspace{7pt}{\large\lineskip .5em
 \begin{tabular}[t]{c}\@name\\[-7pt]\@address
 \end{tabular}\par}\end{center}
 \par\vspace{3pt}}
\makeatother

\makeatletter
\renewcommand{\section}{\@startsection{section}{1}{\z@}%
  {-6pt plus -1pt minus -1pt}{4pt plus .5pt minus .5pt}%
  {\normalfont\normalsize\bfseries}}
\renewcommand{\subsection}{\@startsection{subsection}{2}{\z@}%
  {-5pt plus -1pt minus -1pt}{4pt plus .5pt minus .5pt}%
  {\normalfont\normalsize\bfseries}}
\def\spatex@tabletype{table}
\long\def\@makecaption#1#2{%
  \ifx\@captype\spatex@tabletype\vskip2pt\else\vskip5pt\fi
  \setbox\@tempboxa\hbox{#1. #2}%
  \ifdim\wd\@tempboxa>\hsize
    #1. #2\par
  \else
    \hbox to\hsize{\hfil\box\@tempboxa\hfil}%
  \fi
  \ifx\@captype\spatex@tabletype\vskip5pt\fi}
\makeatother

\begin{document}
\ninept
\setlength{\baselineskip}{9.7pt}
\setlength{\abovedisplayskip}{3pt plus 1pt minus .5pt}
\setlength{\belowdisplayskip}{3pt plus 1pt minus .5pt}
\setlength{\abovedisplayshortskip}{2.5pt plus 1pt minus .5pt}
\setlength{\belowdisplayshortskip}{3pt plus 1pt minus .5pt}
\maketitle

\begin{abstract}
\vspace{-5pt}
Multichannel-to-multichannel target speech extraction (MIMO TSE) aims to recover the target speech signal at each microphone while preserving interchannel spatial information. However, existing MIMO TSE systems generally assume a fixed microphone configuration, limiting their applicability to recording devices with varying microphone configurations. In this work, we propose SPATEX, the first array-flexible MIMO TSE framework that reconstructs microphone-wise target speech signals across microphone counts and array geometries. Given microphone coordinates and a target direction-of-arrival(DOA), SPATEX combines geometry-aware channel exchange and shared temporal attention to jointly reconstruct microphone-wise target signals while maintaining permutation equivariance. Experiments show that SPATEX outperforms array-flexible single-output baselines under most test conditions and consistently improves extraction quality and spatial fidelity over fixed-array MIMO baselines. Ablation studies further validate the benefits of temporal coordination and spatial supervision, while additional evaluations demonstrate robustness to target DOA errors. Our code and dataset are available at \url{https://github.com/Wenanzhi/SPATEX}.

\end{abstract}

\begin{keywords}
target speech extraction, array-flexible signal processing,
spatial information preservation, multichannel-input multichannel-output
\end{keywords}

\begin{figure*}[!t]
    \centering
    \includegraphics[width=\textwidth]{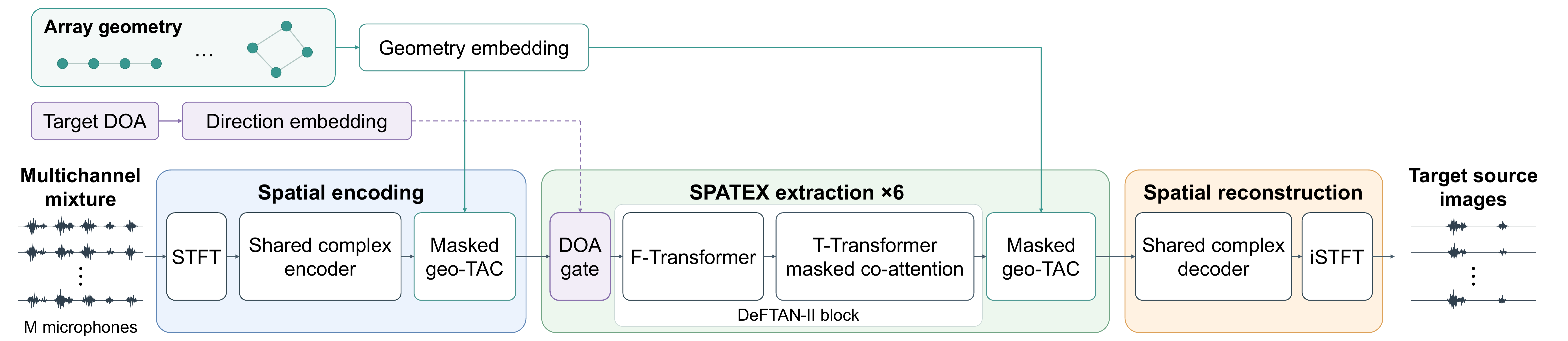}
\caption{\textbf{Overview of SPATEX.} The network reconstructs the target speech image at each microphone using array geometry and target DOA as auxiliary cues. Encoder and decoder parameters are shared across microphones.}
    \label{fig:overview}
\end{figure*}
\section{Introduction}
\label{sec:introduction}
Target speech extraction (TSE) aims to isolate a target speaker from interfering sources using auxiliary cues such as speaker identity\cite{delcroix2020tdspeakerbeam,ge2020spex}, direction~\cite{elminshawi2023beamformer,grondin2020steernet,briegleb2023icospa,tesch2023ssf,jing2025doa,pang2025directional}, or temporal clues. 
For immersive audio applications, including virtual and augmented reality, extracting only the target signal is insufficient, as spatial information is essential for subsequent rendering and interaction. 
Multichannel-to-multichannel target speech extraction (MIMO TSE) addresses this requirement by reconstructing the target signal at each microphone, thereby preserving the interchannel relationships of the target acoustic image. 
However, practical systems typically employ varying numbers of microphones and diverse array geometries, posing challenges for deploying a unified MIMO TSE model across different configurations.

Existing multichannel extraction methods primarily focus on recovering a single-channel target signal by exploiting spatial cues. 
Although these methods effectively utilize microphone information, they discard microphone-wise spatial information in the output. 
MIMO TSE methods address this limitation by reconstructing the target signals at all microphones. 
For example, M2M-TSE and SoundCompass~\cite{choi2025multichannel,choi2026soundcompass} estimate microphone-wise target signals under predefined array geometry, but their array-specific architectures restrict them to fixed microphone numbers and array geometries. 
Meanwhile, array-flexible methods based on microphone permutation-invariant processing, such as TAC-based approaches~\cite{luo2020tac} and FlexIO~\cite{masuyama2026flexio}, support varying input configurations, but typically reconstruct only a reference-channel output rather than the complete target source image. 
These limitations motivate the reconstruction of microphone-wise target spatial source images across varying array configurations.

In this paper, we propose SPATEX (\textbf{S}patial-information \textbf{P}reser\-vation for \textbf{A}rray-flexible \textbf{T}arget speech \textbf{EX}traction), the first array-flexible framework for MIMO TSE. SPATEX employs geometry-aware channel exchange to aggregate information across arbitrary microphone configurations and shared temporal attention to capture microphone-invariant speech characteristics, while retaining microphone-specific representations for spatial reconstruction. 
Conditioned on the target direction of arrival and microphone coordinates, the proposed framework reconstructs the reverberant target image at each available microphone using a single permutation-equivariant model. 
Furthermore, spatial supervision based on interchannel phase difference and coherence explicitly constrains the relationships among reconstructed microphone signals. 
Experimental results demonstrate that SPATEX achieves up to 15.98 dB SI-SNRi on unseen arrays and consistently improves spatial reconstruction accuracy. Compared with fixed-array MIMO baselines, SPATEX reduces interchannel phase and delay errors, achieving 0.614 rad \dipd{} and 5.00 $\mu\mathrm{s}$ \ditd{} on the fixed-array benchmark.
Ablation studies reveal that shared temporal coordination contributes primarily to extraction quality, improving SI-SNRi by up to 3.06 dB, while spatial supervision further reduces interchannel errors and improves direction preservation.
Additional evaluations show that a single SPATEX model generalizes beyond the training microphone range, maintains competitive performance with fixed-array MIMO systems, and remains robust to errors in target direction cues. Our main contributions are as follows:

\begin{itemize}
    \item We introduce SPATEX, the first array-flexible MIMO TSE framework that reconstructs microphone-wise target speech signals across varying microphone counts and array geometries.
    
    \item We design a permutation-equivariant architecture that integrates geometry-aware channel exchange with shared temporal modeling to coordinate cross-microphone processing while preserving microphone-specific spatial characteristics.
    
    \item Extensive experiments demonstrate the effectiveness of SPATEX across diverse array configurations, while ablation studies reveal the complementary roles of temporal coordination and spatial supervision in achieving adaptive extraction and spatially consistent reconstruction.
\end{itemize}

\section{SPATEX for Array-Flexible MIMO TSE}
\label{sec:method}

\subsection{Problem Formulation}
\label{sec:problem}

Consider an $M$-microphone recording of $I$ speech sources in a
reverberant environment:
\begin{equation}
    \mathbf{Y} = \sum_{i=1}^{I}\mathbf{X}_i + \mathbf{N},
    \qquad
    \mathbf{X}_i = s_i * \mathbf{H}_i,
    \label{eq:mixture}
\end{equation}
where $\mathbf{Y},\mathbf{X}_i,\mathbf{N}\in\mathbb{R}^{M\times N}$
denote the mixture, the reverberant image of source $i$, and additive
noise, respectively. The dry source $s_i$ is convolved with its
multichannel room impulse response $\mathbf{H}_i$.

Let $\mathbf{P}=[\mathbf{p}_1,\ldots,\mathbf{p}_M]^\top
\in\mathbb{R}^{M\times 3}$ denote the microphone coordinates in the
array coordinate system. The target azimuth $\theta_g$ specifies the
direction of the target source relative to the array center. Given the
mixture, microphone geometry, and target direction, SPATEX estimates
\begin{equation}
    \widehat{\mathbf{X}}_g
    = f_{\Theta}(\mathbf{Y},\mathbf{P},\theta_g),
    \label{eq:task}
\end{equation}
where $\widehat{\mathbf{X}}_g\in\mathbb{R}^{M\times N}$ contains the
reconstructed target images at all microphones. Unlike conventional
MIMO TSE systems designed for a fixed array, the same parameters
$\Theta$ are shared across varying microphone counts and array
geometries.

\subsection{Model Architecture}
\label{sec:architecture}

As illustrated in Figure~\ref{fig:overview}, SPATEX is designed as an
array-flexible MIMO TSE framework that preserves microphone-specific
spatial information while enabling cross-microphone interaction across
arbitrary array configurations. The framework consists of a shared
encoder, a target extraction module, and a spatial reconstruction
decoder. The encoder and separator operate on microphone-specific
feature streams, while shared transformations facilitate information
exchange without depending on a fixed microphone array.

The input waveforms and microphone coordinates are padded to $K\geq M$
channels with a binary validity mask $b_m$, where
$M=\sum_{m=1}^{K}b_m$. After normalization using a common standard
deviation over valid microphones, a shared encoder transforms the real
and imaginary components of the STFT representation into a
$K\times C\times T\times F$ feature tensor, where $C$, $T$, and $F$
denote the feature channels, frames, and frequency bins, respectively.
The validity mask is propagated throughout the network to prevent padded channels from contributing to feature aggregation.

The extraction module follows the frequency--time modeling design of
DeFTAN-II~\cite{lee2024deftanii} and comprises six separator stages.
Each stage integrates DOA modulation, microphone-wise frequency
modeling, temporal co-attention, and geometry-conditioned
transform--average--concatenate (TAC) operations.
The network parameters are shared across microphones while remaining
independent across stages, enabling a single model to process
different microphone configurations.

For a microphone stream $\mathbf{Z}_m$, TAC first extracts
microphone-specific features and aggregates information from valid
microphones:
\begin{equation}
    \mathbf{h}_m=\mathcal{T}(b_m\mathbf{Z}_m),\quad
    \overline{\mathbf{h}}=\frac{1}{M}\sum_{m=1}^{K}b_m\mathbf{h}_m,
    \label{eq:tac_context}
\end{equation}
where $\mathcal{T}$ is a shared transform. The updated feature is
computed as
\begin{equation}
    \mathbf{Z}'_m=b_m\left\{\mathbf{Z}_m+
    \mathcal{U}([\mathbf{h}_m,\overline{\mathbf{h}},\mathbf{e}_m])
    \right\},
    \label{eq:tac}
\end{equation}
where $\mathcal{U}$ is a shared transform,
$[\cdot]$ denotes concatenation, and $\mathbf{e}_m$ represents the geometry
embedding described in Section~\ref{sec:conditioning}. Since the
aggregation only considers valid microphones, TAC supports varying
microphone counts and maintains permutation equivariance with respect
to microphone ordering.

To model shared speech content across microphones while preserving
microphone-dependent spatial information, SPATEX replaces the temporal
attention operation in DeFTAN-II with channel-invariant
co-attention~\cite{horiguchi2022multichannel,masuyama2026flexio}.
At each frequency bin, microphone-wise queries $\mathbf{Q}_m$,
keys $\mathbf{K}_m$, and values $\mathbf{V}_m$ are derived from
overlapping temporal patches. For each attention head,
\begin{equation}
    \mathbf{A}
    =
    \operatorname{softmax}
    \left(
    \frac{
    \sum_{m=1}^{K}b_m\mathbf{Q}_m\mathbf{K}_m^\top
    }
    {M\sqrt{d_h}}
    \right),
    \quad
    \mathbf{O}_m=\mathbf{A}\mathbf{V}_m ,
    \label{eq:coattention}
\end{equation}
where $d_h$ denotes the query/key dimension and softmax operates over
key-time tokens. The shared attention map captures microphone-invariant
temporal structures, while the microphone-specific values preserve
individual spatial information required for multichannel reconstruction.

The shared decoder estimates the real and imaginary components of the
target STFT for each valid microphone. Inverse STFT and amplitude
rescaling produce the reconstructed target images
$\widehat{\mathbf{X}}_g$. Through masked aggregation and shared transformations, a single model can reconstruct microphone-wise target signals across diverse array configurations.

\subsection{Geometry and Target-Direction Conditioning}
\label{sec:conditioning}

SPATEX employs microphone geometry and target direction as conditioning
signals. The geometry information characterizes microphone-specific
spatial properties, while the target direction specifies the
source to be extracted.

For geometry conditioning, a shared MLP integrates the relative
microphone offset, its normalized representation, and the array radius:
\begin{equation}
    \mathbf{e}_m
    =
    b_m\operatorname{MLP}_{\mathrm{geo}}
    \left(
    [\mathbf{d}_m,\mathbf{d}_m/r,r]
    \right),
    \label{eq:geometry}
\end{equation}
where $\mathbf{d}_m$ denotes the offset from the centroid of valid
microphones and $r$ represents the maximum distance from this centroid.
The normalized offsets characterize the array topology, while the physical offsets
and radius preserve the spatial scale. The geometry embedding is
derived solely from microphone coordinates and reused across all TAC
blocks.

The target azimuth is
represented using the $\ell_2$-normalized cyclic encoding
$c(\theta_g)$. At each separator stage $\ell$, this encoding is
projected through a linear layer, layer normalization, and PReLU to
generate a modulation vector $\mathbf{g}^{(\ell)}$. The stage input is
then modulated as
\begin{equation}
    \widetilde{\mathbf{Z}}_m^{(\ell)}
    =
    \mathbf{Z}_m^{(\ell)}
    \odot
    \mathbf{g}^{(\ell)},
    \label{eq:doa_modulation}
\end{equation}
where $\odot$ denotes element-wise multiplication. The same modulation
vector is applied to all microphone streams, thereby conditioning the
extraction process on the target direction.

\subsection{Spatially Consistent Training Objective}
\label{sec:objective}

To simultaneously optimize target reconstruction and interchannel spatial
consistency, we combine the phase-constrained magnitude (PCM)
loss~\cite{choi2025multichannel} with spatial supervision based on
interchannel phase difference (IPD)~\cite{battula2025phase} and complex coherence~\cite{tolooshams2022stereo}:
\begin{equation}
    \mathcal{L}
    =
    \mathcal{L}_{\mathrm{PCM}}
    +
    \lambda_{\mathrm{IPD}}
    \mathcal{L}_{\mathrm{IPD}}
    +
    \lambda_{\mathrm{coh}}
    \mathcal{L}_{\mathrm{coh}} .
    \label{eq:objective}
\end{equation}

PCM supervises target reconstruction at individual microphones,
whereas IPD and coherence losses explicitly constrain the spatial
relationships among reconstructed signals. 
Spatial losses are computed over target-active TF bins and valid unordered microphone
pairs within each scene, and then averaged across scenes.

\section{Experimental Setup}
\label{sec:experiments}

We evaluate SPATEX from three perspectives:
(1) extraction quality and spatial fidelity under varying microphone
configurations, (2) generalization to unseen microphone counts and
array geometries, and (3) comparison with existing fixed-array
MIMO TSE systems.

\subsection{Dataset and Evaluation Protocol}

Reverberant mixtures are generated from the speaker-disjoint LibriSpeech~\cite{panayotov2015librispeech} subsets train-clean-100, dev-clean, and test-clean for training, validation, and testing, respectively.
The reverberant target image at each microphone serves as
the reconstruction reference. Each 8-kHz mixture contains one
target speaker and two interfering speakers. Room impulse responses are
generated using the image-source method in
pyroomacoustics~\cite{scheibler2018pyroomacoustics} with randomly
sampled rooms, source positions, and array configurations. We sample
$T_{60}$ from 0.15--0.7~s, enforce at least $15^\circ$ angular
separation between sources, and sample each interferer's input SIR from
$[-5,5]$~dB. No additional background noise is added. 

Variable-array evaluation covers two geometry conditions (Matched Geometry and Unmatched Geometry) and two micro\-phone-count ranges (2–8 and 9–10).
The geometry conditions are defined relative to the array shapes used
during training.

\begin{itemize}

\item \textbf{Matched Geometry:}
We use uniform linear arrays and regular planar grid arrays with
2.5--8~cm spacing between adjacent microphones and no position
perturbations. Variable-array training uses these array types with
2--8 microphones.

\item \textbf{Unmatched Geometry:}
Sparse arrays are formed by selecting noncontiguous elements from a
planar grid. Random arrays are generated with apertures of 5--25~cm.
Aperture denotes the maximum distance between any microphone pair.
Both array types undergo position perturbations with a standard deviation
of 1~cm. The perturbed coordinates are used for both RIR simulation and
model conditioning.

\end{itemize}

Each geometry condition contains 500 mixtures with 2--8 microphones and
1,000 mixtures with 9--10 microphones, equally divided between nine and
ten microphones. 
The 9--10-microphone subsets evaluate microphone-count
extrapolation. The ablation study in Table~\ref{tab:variable_joint} uses
Matched Geometry with 2--8 microphones and Unmatched Geometry with
9--10 microphones.

The Fixed 4-Mic Array condition uses a four-microphone linear array with
4-cm adjacent microphone spacing. The test set contains 500 mixtures.

\subsection{Training Details}

All learned baselines are retrained on the corresponding variable-array
or fixed-array training data. SPATEX (variable-array) is trained on the
variable-array data and used unless otherwise specified. SPATEX (fixed-4)
is trained separately on the fixed-array data. SPATEX uses a 256-sample Hann-windowed STFT
with a 128-sample hop and 64-dimensional features.
In~\eqref{eq:objective}, we set $\lambda_{\rm IPD}=0.15$ and
$\lambda_{\rm coh}=0.03$. SPATEX is optimized using Adam with an initial
learning rate of $5\times10^{-4}$ and batch size 8 for at most 100 epochs.
Models are selected according to validation SI-SNRi.

\begin{table*}[t]
 \centering
 \caption{\textbf{Effect of temporal coordination and spatial supervision on target extraction and spatial fidelity.}
Independent and Shared denote microphone-specific and shared temporal
attention, respectively. All variants use PCM; IPD and Coh denote
interchannel phase difference and coherence losses. Pres@5 denotes the
percentage of mixtures with output--target DOA drift at most $5^\circ$.}
 \label{tab:variable_joint}
 \scriptsize
 \setlength{\tabcolsep}{1pt}
 \renewcommand{\arraystretch}{0.96}
 \def\cfgcell#1{\makebox[2em][c]{#1}}
\resizebox{\textwidth}{!}{%

\begin{tabular}{@{}ll@{\hspace{6pt}}c@{\hspace{4pt}}c@{\hspace{6pt}}cccc@{\hspace{8pt}}ccc@{\hspace{8pt}}cc@{}}
  \toprule
  \multirow{2}{*}[-0.5ex]{Testset} & \multicolumn{1}{@{}c@{\hspace{5.5pt}}}{Temporal} &
  \multicolumn{2}{@{\hspace{1.5pt}}c@{\hspace{3pt}}}{Spatial losses} &
  \multicolumn{4}{@{}c@{\hspace{7.5pt}}}{Signal extraction} &
  \multicolumn{3}{@{}c@{\hspace{7.5pt}}}{Pairwise spatial fidelity} &
  \multicolumn{2}{@{}c@{}}{Direction preservation} \\
  \cmidrule(l{0pt}r{5.5pt}){2-2}\cmidrule(l{1.5pt}r{3pt}){3-4}\cmidrule(l{0pt}r{7.5pt}){5-8}
  \cmidrule(l{0pt}r{7.5pt}){9-11}\cmidrule(l{0pt}r{0pt}){12-13}
  & Affinity & \cfgcell{IPD} & \cfgcell{Coh} &
  \sisnri{} $\uparrow$ & \snri{} $\uparrow$ & PESQ $\uparrow$ &
  STOI $\uparrow$ & \dipd{} $\downarrow$ & \ditd{} $\downarrow$ &
  \dild{} $\downarrow$ & \doadrift{} $\downarrow$ &
  \presfive{} $\uparrow$ \\
  \midrule
  \multirow{5}{*}{\shortstack[l]{Matched Geometry\\2--8 microphones}}
  & Independent & \cfgcell{$\checkmark$} & \cfgcell{$\checkmark$} &
  11.31 & 12.19 & 2.89 & 0.831 & 0.728 & 8.95 & 0.182 & 3.334 & 84.6 \\
  & Shared & \cfgcell{--} & \cfgcell{--} &
  11.03 & 11.98 & 2.88 & 0.821 & 0.801 & 14.42 & 0.187 & 9.807 & 57.0 \\
  & Shared & \cfgcell{$\checkmark$} & \cfgcell{--} &
  13.79 & 14.40 & 3.30 & 0.879 & 0.636 & 9.11 & 0.166 & 3.863 & 84.2 \\
  & Shared & \cfgcell{--} & \cfgcell{$\checkmark$} &
  13.13 & 13.81 & 3.20 & 0.865 & 0.679 & 8.55 & \textbf{0.149} & 2.965 & 87.4 \\
  \cmidrule(l{1pt}r{1pt}){2-13}
  & \textbf{Shared (SPATEX)} &
  \cfgcell{$\checkmark$} & \cfgcell{$\checkmark$} &
  \textbf{14.23} & \textbf{14.81} & \textbf{3.41} &
  \textbf{0.891} & \textbf{0.615} & \textbf{6.00} &
  0.165 & \textbf{2.871} & \textbf{90.4} \\
  \midrule
  \multirow{5}{*}{\shortstack[l]{Unmatched Geometry\\9--10 microphones}}
  & Independent & \cfgcell{$\checkmark$} & \cfgcell{$\checkmark$} &
  12.35 & 12.92 & 3.05 & 0.858 & 0.774 & 11.24 & 0.228 & 2.150 & 93.4 \\
  & Shared & \cfgcell{--} & \cfgcell{--} &
  12.18 & 12.77 & 3.04 & 0.858 & 0.829 & 12.98 & 0.218 & 6.430 & 78.4 \\
  & Shared & \cfgcell{$\checkmark$} & \cfgcell{--} &
  15.41 & 15.75 & 3.54 & 0.912 & 0.635 & 7.94 & 0.209 & 1.485 & 95.9 \\
  & Shared & \cfgcell{--} & \cfgcell{$\checkmark$} &
  14.65 & 15.06 & 3.41 & 0.900 & 0.677 & 7.77 & \textbf{0.187} & 1.315 & 96.7 \\
  \cmidrule(l{1pt}r{1pt}){2-13}
  & \textbf{Shared (SPATEX)} &
  \cfgcell{$\checkmark$} & \cfgcell{$\checkmark$} &
  \textbf{15.98} & \textbf{16.26} & \textbf{3.65} &
  \textbf{0.923} & \textbf{0.604} & \textbf{6.89} &
  0.203 & \textbf{1.235} & \textbf{97.0} \\
  \bottomrule
 \end{tabular}%
 }
\end{table*}

\begin{table}[ht]
 \centering
 \caption{\textbf{Extraction performance of SPATEX and array-flexible single-output baselines.}
Results are reported as reference-channel \sisnri{} (dB, $\uparrow$).
Matched and Unmatched denote array shapes seen and unseen during
training, respectively. Tests with 9--10 microphones evaluate
extrapolation beyond the training range of 2--8 microphones.}
 \label{tab:variable_signal}
 \footnotesize
 \setlength{\tabcolsep}{2pt}
 \renewcommand{\arraystretch}{0.95}
 \begin{tabular*}{\columnwidth}{@{\extracolsep{\fill}}lccccc@{}}
  \toprule
  \multirow{2}{*}[-0.5ex]{System} & \multirow{2}{*}[-0.5ex]{Params. (M)} & \multicolumn{2}{c}{2--8 microphones} & \multicolumn{2}{c}{9--10 microphones} \\
  \cmidrule(r{2pt}){3-4}\cmidrule(l{2pt}){5-6}
  & & Matched & Unmatched & Matched & Unmatched \\
  \midrule
  Analytic DSB & -- & -0.55 & -0.39 & -1.10 & -0.39 \\
  SteerNet-GEV & 0.83 & 1.14 & 1.22 & 0.51 & 1.11 \\
  \shortstack[l]{D-SE-BSRNN} & 2.70 & 5.81 & 6.18 & 6.13 & 6.80 \\
  BG-TSE & 6.56 & 5.88 & 6.01 & 6.37 & 6.52 \\
  USES-DOA & 3.12 & 12.53 & 13.01 & 13.18 & 13.90 \\
  FlexIO-DOA & 7.36 & 13.91 & \textbf{14.57} & 14.60 & 15.39 \\
  \midrule
  \textbf{SPATEX} & 4.57 & \textbf{14.23} & 13.82 & \textbf{15.04} & \textbf{15.98} \\
  \bottomrule
 \end{tabular*}
\end{table}

\begin{table}[htbp]
 \centering
 \caption{\textbf{Extraction quality and spatial fidelity on the Fixed 4-Mic Array.}
Variable-array and fixed-4 denote the SPATEX training configurations.
+IPD+Coh adds SPATEX's phase and coherence losses to M2M-TSE-DOA.}
 \label{tab:fixed_compat}
 \setlength{\tabcolsep}{2pt}
 \renewcommand{\arraystretch}{1.08}
 \resizebox{\columnwidth}{!}{%
 \begin{tabular}{@{}lcccc@{}}
  \toprule
  System & Params. (M) & \sisnri{} $\uparrow$ &
  \dipd{} $\downarrow$ & \ditd{} $\downarrow$ \\
  \midrule
  Waveformer-M2M & 9.10 & 1.57 & 0.899 & 24.29 \\
  DeFTAN-II & 4.03 & 9.69 & 0.815 & 10.58 \\
  M2M-TSE-DOA & 4.03 & 10.72 & 0.761 & 7.67 \\
  M2M-TSE-DOA + IPD + Coh & 4.03 & 10.43 & 0.696 & 7.17 \\
  SoundCompass & 2.54 & 11.68 & 0.773 & 8.04 \\
  \midrule
  \textbf{SPATEX (variable-array)} & 4.57 & 12.90 & \textbf{0.614} & 5.00 \\
  \textbf{SPATEX (fixed-4)} & 4.57 & \textbf{13.22} & 0.627 & \textbf{4.75} \\
  \bottomrule
 \end{tabular}
 }
\end{table}

\subsection{Evaluation Metrics}

Extraction quality is evaluated at reference microphone 1 with respect
to its
reverberant target image using \sisnri{}, \snri{}~\cite{leroux2019sdr},
PESQ~\cite{rix2001pesq}, and STOI~\cite{taal2011stoi}. \sisnri{} and
\snri{} measure improvements over the mixture in dB, while PESQ is computed
in narrowband mode.

For MIMO outputs, spatial fidelity is assessed using the mean absolute
errors of interchannel cues between the reconstructed and target images.
\dild{} (dB) is computed from whole-utterance energy ratios, \dipd{} (rad) uses
circular phase differences, and \ditd{} ($\mu$s) is estimated using GCC-PHAT with a
125-$\mu$s lag grid~\cite{choi2025multichannel,knapp1976gcc}.
Each error is averaged over valid microphone pairs within each mixture
and then across mixtures.

To evaluate direction preservation, we estimate directions from the
reconstructed and target images using the same SRP-PHAT
configuration~\cite{dibiase2001robust}. DOA drift is defined as the circular
difference between these estimates, accounting for front/back ambiguity
in linear arrays. We report the mean drift in degrees and \presfive{},
the percentage of mixtures with drift no greater than $5^\circ$.

\section{Results and Analysis}
\label{sec:results}

\subsection{Analysis of Temporal Coordination and Spatial Supervision}
Table~\ref{tab:variable_joint} evaluates temporal affinity and spatial losses under Matched Geometry and Unmatched Geometry. Independent computes temporal attention separately for each microphone, whereas Shared averages attention logits across valid microphones before softmax and applies the resulting attention matrix to the value features of each microphone. 
Under the same PCM+IPD+Coh objective, the shared affinity mechanism improves \sisnri{} by 2.92 and 3.63~dB, respectively, while enhancing all reported spatial metrics. These improvements generalize to nine and ten microphones with array configurations unseen during training.

Adding either spatial loss to PCM improves extraction performance and spatial fidelity. Among the individual spatial losses, IPD achieves better extraction and phase accuracy, whereas coherence loss reduces level and delay errors. The combined objective achieves the best extraction quality, the lowest phase and delay errors, and the best direction preservation on both test sets. PCM+Coh yields the lowest \dild{}.

\subsection{Comparison with Array-Flexible Single-Output Baselines}
Existing MIMO TSE methods assume fixed microphone configurations. Therefore, we compare SPATEX with array-flexible single-output baselines to evaluate extraction performance across microphone counts and array geometries.

Baselines include training-free Analytic DSB~\cite{vanveen1988beamforming}, SteerNet-GEV~\cite{grondin2020steernet}, D-SE-BSRNN~\cite{gu2024rezero}, BG-TSE~\cite{elminshawi2023beamformer}, USES-DOA~\cite{zhang2023uses}, and FlexIO-DOA~\cite{masuyama2026flexio}. We adapt D-SE-BSRNN by applying shared IPD encoding within each frequency band, followed by mean pooling over both pair orientations and all valid microphone pairs. USES-DOA incorporates direction and geometry embeddings to its encoder features, while FlexIO-DOA introduces direction and geometry conditioning into its speech prompt. Both methods retain their original backbones and are trained for 100 epochs using negative SNR loss on the reverberant target image of the reference channel.

Table~\ref{tab:variable_signal} shows that SPATEX achieves the highest \sisnri{} in three of the four conditions, including both 9--10-microphone subsets. These results demonstrate that SPATEX provides competitive extraction performance compared with array-flexible single-output systems while reconstructing microphone-wise target images using a single model.

\subsection{MIMO Reconstruction on a Fixed Array}
\label{sec:fixed_mimo}

Existing MIMO TSE baselines assume fixed microphone configurations. Therefore, we compare SPATEX with these systems on the Fixed 4-Mic Array to evaluate the spatial fidelity of microphone-wise target reconstruction. SPATEX (fixed-4) is trained on the same fixed-array
data as the baselines for a fair comparison, whereas SPATEX (variable-array) is trained on variable-array data and evaluated directly without array-specific fine-tuning.

Waveformer-M2M adapts Waveformer~\cite{veluri2023waveformer} to four-channel input and output with target-DOA conditioning. DeFTAN-II~\cite{lee2024deftanii} adopts a DOA-conditioned MIMO configuration. M2M-TSE-DOA~\cite{choi2025multichannel} retains the original direction-only configuration with a time-invariant DOA embedding. Its spatially supervised variant incorporates SPATEX's IPD and coherence losses with identical weights. SoundCompass~\cite{choi2026soundcompass} operates at 8~kHz using a single stage, sums the direct-path and reverberation outputs, and applies a shared signed least-squares gain estimated from the mixture.

As shown in Table~\ref{tab:fixed_compat}, both SPATEX variants achieve superior extraction quality and lower spatial errors than the baselines. 
Incorporating IPD and coherence losses to M2M-TSE-DOA reduces phase and delay errors, with a slight degradation in extraction quality. Both SPATEX variants outperform this spatially supervised baseline in extraction quality and spatial fidelity.

\subsection{Robustness to DOA Cue Errors}

\begin{figure}[!htbp]
 \centering
 \includegraphics[width=\columnwidth]{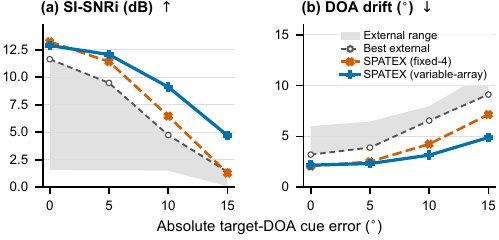}
 \caption{\textbf{Robustness to target-DOA cue errors on the Fixed 4-Mic Array.}
Results are averaged over positive and negative offsets of equal magnitude.
Shaded regions show the range of the four external baselines; gray dashed
curves indicate their best performance at each error magnitude.}
 \label{fig:doa}
\end{figure}

Figure~\ref{fig:doa} shows that both SPATEX variants outperform all baselines in Table~\ref{tab:fixed_compat} in terms of extraction quality and direction preservation for cue errors up to $10^\circ$. As the cue error increases, variable-array SPATEX exhibits a smaller performance degradation than fixed-4 SPATEX on both metrics. The increasing performance gap indicates improved robustness to directional mismatch.

\section{Conclusion}

In this paper, we propose SPATEX, the first array-flexible MIMO TSE framework for microphone-wise target speech image reconstruction. Geometry-aware channel exchange and shared temporal attention enable coordinated reconstruction across microphones, while spatial supervision preserves interchannel relationships. Experiments demonstrate that SPATEX outperforms array-flexible single-output baselines under most test conditions and improves both extraction quality and spatial fidelity over fixed-array MIMO systems. The variable-array model generalizes beyond the training microphone range without fine-tuning and remains robust to target DOA cue errors.

\pagebreak[4]
\section{Compliance with Ethical Standards}

This study uses simulated reverberant mixtures generated from the
publicly available LibriSpeech corpus
for model training and evaluation. No new participants were recruited
or recorded, and no animal experiments were conducted. The corpus
was used in accordance with its licensing terms. No additional
institutional ethical approval was required for the reported experiments.
\section{Acknowledgments}
The authors declare no actual or potential conflicts of interest.
\bibfont
\bibliographystyle{IEEEbib}
\bibliography{references}

\end{document}